\documentclass[useAMS]{mn2e}
\usepackage{graphicx}	% Including figure files

\usepackage{times}
\usepackage{amssymb,amsmath}

\def\lsim{ \lower .75ex\hbox{$\sim$} \llap{\raise .27ex \hbox{$<$}} }
\def\gsim{ \lower .75ex \hbox{$\sim$} \llap{\raise .27ex \hbox{$>$}} }

\title[Neutrinos from FR0] 
{High-energy neutrinos from  FR0 radio-galaxies?}

\author[Tavecchio et al.]
{F. Tavecchio$^1$\thanks{E--mail: fabrizio.tavecchio@brera.inaf.it}, C. Righi$^{2,1}$, A. Capetti$^3$, P. Grandi$^4$, G. Ghisellini$^1$ 
\\
$^1$ INAF -- Osservatorio Astronomico di Brera, via E. Bianchi 46, I--23807 Merate, Italy\\
$^2$ Universit\`a degli Studi dell'Insubria, Via Valleggio 11, I-22100 Como, Italy\\
$^3$ INAF -- Osservatorio Astronomico di Torino, via Osservatorio 20, 10025 Pino Torinese, Italy,\\
$^3$ INAF--IASFBO, Via Gobetti 101, I-40129 Bologna, Italy}

\begin{document}

% \date{Accepted 1988 December 15. Received 1988 December 14; 
% in original form 1988 October 11}

%\pagerange{\pageref{firstpage}--\pageref{lastpage}} \pubyear{2007}

\maketitle

\begin{abstract} 
The sources responsible for the emission of high-energy ($\gtrsim$ 100 TeV) neutrinos detected by IceCube are still unknown. Among the possible candidates, active galactic nuclei with relativistic jets are often examined, since the outflowing plasma seems to offer the ideal environment to accelerate the required parent high-energy cosmic rays. The non-detection of single point sources or -- almost equivalently -- the absence, in the IceCube events, of multiplets originating from the same sky position, constrains the cosmic density and the neutrino output of these sources, pointing to a numerous population of  faint  sources.  Here we explore the possibility that {\it FR0 radiogalaxies}, the population of compact sources recently identified in large radio and optical surveys and representing the bulk of radio-loud AGN population, can represent suitable candidates for neutrino emission. Modeling the spectral energy distribution of a FR0 radiogalaxy recently associated to a $\gamma$-ray source detected by the Large Area Telescope onboard {\it Fermi}, we derive the physical parameters of its jet, in particular the power carried by it. We consider the possible mechanisms of neutrino production, concluding that $p\gamma$ reactions in the jet between protons and ambient radiation is too inefficient to sustain the required output. We propose an alternative scenario, in which protons, accelerated in the jet, escape from it and diffuse in the host galaxy, producing neutrinos as a result of $pp$ scattering with the interstellar gas, in strict analogy with the processes taking place in star-forming galaxies.
\end{abstract}

\begin{keywords} astroparticle physics --- neutrinos --- galaxies: jets -- radiation mechanisms: non-thermal ---  $\gamma$--rays: galaxies %-- galaxies: general
\end{keywords}

\section{Introduction}

The astrophysical sources responsible for the extraterrestrial (i.e. in excess of the atmospheric) neutrino flux detected by IceCube at PeV energies (Aartsen et al. 2013, 2014) are still unknown.  The events (more than $\sim$ 50 above 100 TeV) display a distribution on the sky consistent with isotropy (with a non significant  excess towards the galactic center) suggesting an extragalactic origin, although there are hints on a possible contribution from a galactic component at low energies (e.g. Ahlers \& Murase 2014, Neronov \& Semikoz 2015, Aartsen et al. 2015, Palladino \& Vissani 2016).  The possible extragalactic astrophysical sources include propagating comic rays (e.g., Kalashev et al. 2013), star-forming and starburst galaxies (e.g., Loeb \& Waxman 2006, Tamborra et al. 2014), AGN-driven shocks (e.g. Lamastra et al. 2016), galaxy clusters (e.g., Murase et al. 2008, Zandanel et al. 2014), $\gamma$--ray burst (e.g., Waxman \& Bahcall 1997) and active galactic nuclei (AGN, e.g., Stecker \& Salamon 1996, Mannheim 1995, Atoyan \& Dermer 2003).

In the case of AGN, it is widely presumed that those hosting relativistic jets could offer the most suitable environment for the production of the high-energy neutrinos. Indeed relativistic jets seems to provide the ideal conditions to accelerate hadrons to the energy ($E_{\rm p}\approx 10^{16-17}$ eV) required to produce neutrinos via the production of pions and their prompt decay ($\pi ^{\pm}\to \mu^{\pm}+\nu_{\mu}\to e^{\pm} + 2\nu_{\mu} +\nu_{\rm e}$; in this paper we do not distinguish among $\nu$ and $\bar{\nu}$). Both blazars (jetted AGN whose outflow point toward the Earth) and radiogalaxies have been discussed as possible sources for the neutrino flux (e.g., Mannheim 1995, Atoyan \& Dermer 2003, Murase et al. 2014, Tavecchio et al. 2014, Padovani et al. 2016, Kadler et al. 2016, Righi et al. 2017, Hooper 2016). In this context, the recent detection of a muon neutrino event whose reconstructed incoming direction is spatially consistent with the position of the gamma-ray flaring BL Lac TXS 0506+056 (Kopper \& Blaufuss 2017, Tanaka et al. 2017) has attracted great attention.

Murase \& Waxman (2016) pointed out that the absence of {\it multiplets} among IceCube events (i.e. the absence of two or more events with the same reconstructed direction) put strong constraints on the characteristics of the population responsible for the bulk of the observed flux. In this way they concluded that several of the proposed source classes can be already ruled out by the present data (see also Palladino \& Vissani 2017 for a similar analysis focused on BL Lac). The idea at the base is that, in order to account for the observed neutrino sky intensity, {\it rare} (i.e. with a too low spatial density) sources must be {\it bright} neutrino emitters and thus they should already pop-up in the data as sources of multiplets. In order to decrease the probability to produce multiple events, suitable candidates should be represented by abundant and (relatively) faint sources (as, e.g., star forming galaxies, Loeb \& Waxman 2006; but see Murase et al. 2016 and Bechtol et al. 2017).
\\

Radiogalaxies represent the larger population of radio-loud AGNs. Their phenomenology is dominated by the presence of a relativistic jets outflowing from the central regions close to a supermassive black hole. Based on their radio power, radiogalaxies are classically divided (after Fanaroff \& Riley 1974) in Fanaroff-Riley type II (FRII, high power) and type I (FRI low power). This  classification corresponds also to a clear difference in the morphology of the radio-emission from the jet. FRII jets tend to be long (several hundreds of kpc, up to 1 Mpc) and collimated, ending with a bright hot spot feeding a radio lobe. On the contrary, the jet associated to FRI seems to be prone to instabilities, which determine a rapid deceleration of the jet (e.g., Bodo et al. 2013), producing edge-darkened plumes and tails.

The cross-matching of large area radio and optical surveys has recently revealed, among the radio-loud AGNs, the existence of an extremely numerous population of very weak sources (Baldi et al. 2015, 2017). For several aspects these objects appear to belong to the continuation at low powers of the classical FRII and FRI radio-galaxies, hence the proposed name ``FR0" (Ghisellini 2011). However, FR0 also show some peculiarities, the most striking being the lack of {\it extended} (i.e. kpc scale) radio-emission, suggesting that the jet -- whose existence is flagged by the relatively bright non-thermal emission of the core -- is not able to reach the large scales as observed in the case of both FRI and FRII radiogalaxies. Quite interestingly, one of these FR0 galaxies has been recently associated to a $\gamma$-ray source detected by LAT (Grandi et al. 2016), demonstrating that the small-scale jet is likely very similar (although with less power) to that of the classical radiogalaxies. It is also clear that the known FR0 are just the tip of the iceberg of a quite abundant population (Baldi et al. 2017). 

In this Letter we intend to explore the possibility that the population of FR0 active galaxies has the right properties to be the long-sought neutrino emitters (). In Sect. 2 we describe a new sample of FR0 that can be used to obtain for the first time the radio flux distribution  of these sources. In Sect. 3 we consider the spectral energy distribution of the benchmark FR0 Tol 1326-379 detected by LAT and we model it,  deriving the basic physical parameters of its jet. These are used in Sect. 4, where we discuss the conditions for the FR0 to be suitable candidate neutrino sources.

Throughout the paper, the following cosmological  parameters are assumed: $H_0=70$ km s$^{-1}$ Mpc$^{-1}$, $\Omega_{\rm M}=0.3$, $\Omega_{\Lambda}=0.7$.  We  use the notation $Q=Q_X \, 10^X $ in cgs units.

\section{The flux distribution of FR0}
\label{sample}

\begin{figure}
\vspace*{-1.5 truecm}
\hspace*{-1.7 truecm}
\includegraphics[width=11.0cm]{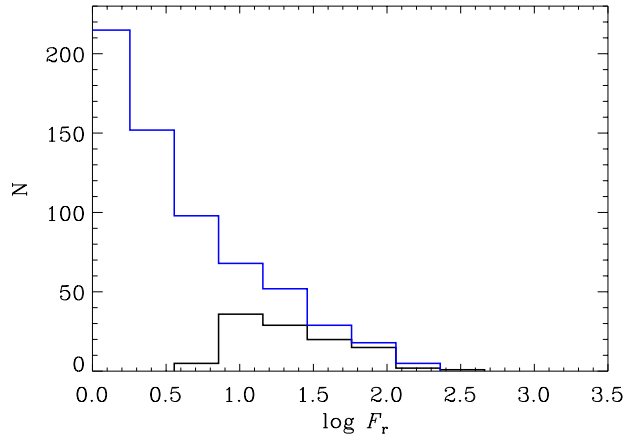}
\vspace*{-6.9truecm}
\caption{Distribution of the radio fluxes (in mJy) of the 108 FR0 (black) and of the sample of 638 elliptical compact radio sources (blue).}
\label{hist}
\end{figure}

We considered the sample of FR0s selected by Baldi et al. (2017). They
searched for compact radio AGN in the sample built by Best \& Heckman (2012) by
cross-matching data from Sloan Digital Sky Survey (Abazajian et al. 2009),
the Faint Images of the Radio Sky at Twenty centimeters survey,
(Becker et al. 1995, Helfand et al. 2015), and the National Radio Astronomy Observatory Very
Large Array Sky Survey (Condon et al. 1998). Baldi et al. extracted 108 FR0
sources requiring a redshift $z \leq 0.05$ and a deconvolved size smaller than
4\arcsec, corresponding to a radius of $<$ 2.5 kpc. Their radio flux
distribution (see Fig. \ref{hist}) spans from the sample limit (5 mJy) to
$\sim$ 400 mJy and peaks at $\sim$20-30 mJy. The presence of such peak
indicates that the selected sample is incomplete at a flux higher than the
selection threshold.

In order to extend the FR0 population to lower fluxes, we consider the 638
elliptical galaxies with $z<0.05$ present in the FIRST source catalog, limited
to a threshold of 1 mJy. Although we cannot obtain an accurate size
measurement for many of these fainter radio galaxies, most of them are
consistent with being unresolved sources in the survey images. The derived flux distribution (LogN-LogS) can be approximately characterized by a power law with slope 1.7, i.e. $N(F_{\rm r})\propto F_{\rm r}^{-1.7}$.

\section{Tol 1326-379: a benchmark FR0}

\begin{figure}
%\hskip -2 cm
\vspace*{-1.7 truecm}
%\vspace*{-1.5 truecm}
\hspace*{-0.2 truecm}
%\vskip -0.3 cm
\includegraphics[width=9cm, height=10.cm]{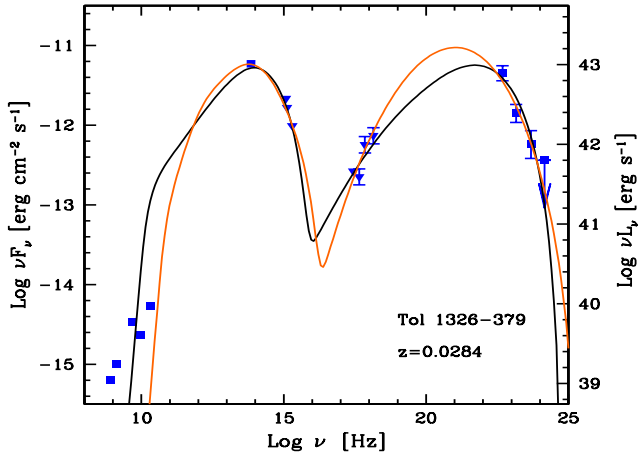} 
\vspace{-2.5 cm}
\caption{The SED of the FR0 radiogalaxy Tol 1326-379 (from Grandi et al. 2016 and Torresi et al. in prep).  The solid lines show the emission calculated with a leptonic one-zone model for the jet (with the parameters reported in Table 1) for the blazar (orange) and the radiogalaxy case (black), respectively. See text for details.
}
\label{sed}
\end{figure}

Grandi et al. (2016)  associate the FR0 radiogalaxy Tol 1326-379 to a $\gamma$-ray source detected by LAT (3FGLJ1330.0--818). Its spectral energy distribution (SED), assembled with the available data, is shown in Fig.\ref{sed}. The ``double-humped" shape closely resemble that displayed by other jet-dominated radio-loud AGNs, blazars and radiogalaxies, for which the bulk of the non-thermal emission is thought to be produced in the part of the jet close to the central engine. This evidence strongly supports the view that, at small scales ($\lesssim$ few pc), the  jet hosted by FR0 is similar to those residing in the classical radio-loud objects.
As discussed in Grandi et al. (2016), the data points in the optical-UV band are likely contaminated by optical emission from the galaxy and should be formally treated as upper limits to the jet emission. On the other hand, the IR datapoint (from {\it WISE}) is probably associated to the non-thermal jet emission. 

% ---------------------------------------------------
% ---------------------------------------------------

To derive the basic physical quantities of the jet, we model the SED of Tol 1326-379 using a standard leptonic one-zone model (Maraschi \& Tavecchio 2003). In brief, we assume that the observed non-thermal continuum is produced through synchrotron and synchrotron self Compton emission by a population of relativistic electrons filling a spherical source of radius $R$, moving with bulk Lorentz factor $\Gamma$ at a viewing angle $\theta _{\rm v}$ with respect to the line of sight. The electrons follow, between $\gamma_{\rm min}$ and $\gamma_{\rm max}$, a smoothed broken power law energy distribution parametrized by the slopes $n_1$ and $n_2$ below and above the Lorentz factor $\gamma_{\rm b}$. The emission produced in the region is then boosted in the observer frame.

The degree of boosting is dictated by the relativistic Doppler factor, $\delta=[\Gamma(1-\beta \cos \theta_{\rm v})]^{-1}$, which critically depends on the observing angle and on the bulk Lorentz factor of the flow. Both quantities are not known for the specific case under study (although the available observational evidence is in favor of a low degree of beaming). We model the SED in two different scenarios, namely assuming 1) that the jet is closely aligned toward the observer (as for the case of blazars) and has a large Lorentz factor ({\it blazar} case) or 2) that the jet is misaligned and slow ({\it radiogalaxy} case). In the latter case we assume a representative viewing angle of $\theta_{\rm v}=30$ degrees. These two cases correspond to the situation commonly assumed for blazars and radiogalaxies. For the latter a possibility explaining the lower value of the Lorentz factor is that the jet in radio-loud AGN is stratified, with a fast ($\Gamma\approx 10$) core surrounded by a slower layer (e.g. Ghisellini et al. 2005). For blazars the viewing angle is small and the boosted spine emission dominates, while for large  angles ($\theta_{\rm v}>1/\Gamma$) the spine emission is deboosted and the observed emission is dominated by the less beamed emission from the slower layer.

The parameters found reproducing the SED with the two cases are reported in Table 1 and the corresponding SEDs are shown in Fig. \ref{sed}. In the Table we also report the energy flux (or power) carried by the jet, in the form of relativistic electrons, $P_{\rm e}$, magnetic field, $P_{B}$, and (cold) protons, $P_{\rm p}$. In both cases the derived total energy flux is of the order of, $P_{\rm j}\simeq {\rm few }\, \times 10^{44}$ erg s$^{-1}$. These values of the power corresponds to the low power tail of the BL Lac population as derived by Ghisellini et al. (2010), consistently with the view that FR0 radiogalaxies represent the low-power extension of the classical radiogalaxy population (and possibly the parent population of the weakest BL Lacs). We note that the derived jet power could be underestimeted in the (likely) case in which the ratio of the cold protons over the relativistic electrons is larger than unity. Finally we also note that a value of the jet power consistent with the SED modeling, $P_{\rm j}\approx 10^{44}$ erg s$^{-1}$, can be derived using the method outlined by Heinz, Merloni \& Schwab (2007).

% -----------------------------------------------------------
\begin{table*} 
\begin{center}
\begin{tabular}{clllllllllllllll|}
\hline
\hline
$R$       
& $B$  
& $\gamma_{\rm min} $  
& $\gamma_{\rm b} $ 
& $\gamma_{\rm max}$  
& $n_1$
& $n_2$
& $\Gamma$ 
& $\theta_{\rm v}$ 
& $P_{\rm e}$
& $P_{\rm B}$
& $P_{\rm p}$
& $P_{\rm j}$
\\
cm  &G &  & & &  & &  &deg. & erg s$^{-1}$ &erg s$^{-1}$ &erg s$^{-1}$ &erg s$^{-1}$\\
\hline  
$3.5\times 10^{15}$ &0.1  &$400$ &$5\times 10^3$ &$3\times10^{4}$ &2  &4.8   &10  &5  &$1.2\times 10^{44}$ & $5\times 10^{40}$& $1.5\times 10^{44}$&$2.7\times 10^{44}$\\ 
$5\times 10^{16}$ &0.2  &$100$ &$ 10^4$ &$3\times10^{4}$ &2  &4.8   &2  &30 &$3.5\times 10^{43}$ & $1.5\times 10^{42}$& $1.2\times 10^{44}$&$1.6\times 10^{44}$\\ 
\hline
\hline
\end{tabular}                                                         
\caption{Input parameters of the models shown in Fig. \ref{sed}. All quantities (except the bulk Lorentz
factors $\Gamma$, the viewing angle $\theta_{\rm v}$ and the powers) are measured
in the rest frame of the emitting plasma.}
\end{center}
\label{tab1}
\end{table*}                                                                  
% --------------------------------------------------------------

In the following we assume these figures derived for Tol 1326-379 as a benchmark to discuss the implications for the production of neutrinos. As a {\it caveat}, we note that the scope of this exercise is necessarily limited, since the properties of the FR0 are far from clear and we are not sure that Tol 1326-379 can be really considered a representative case for the overall FR0 population. An analysis similar to that performed for Tol 1326-379 extended to other FR0 could be useful to better assess the properties of these sources. Another remark concerns the fact that, although we considered the cases in which the jet is either aligned (blazar case) or not not-aligned (radiogalaxy) to the line of sight, for the following discussion we assume the latter configuration. In any case, as far as the jet power is concerned, both cases are very similar.

\section{FR0 as neutrino emitters}

Having estimated some of the basic quantities characterizing the FR0 population and their jets, we explore here the possibility that FR0 are emitters of high-energy neutrinos.

A first important consequence can be drawn from the distribution of the radio fluxes shown in Fig. \ref{hist}. As a working hypothesis we assume that neutrino and radio outputs are linked by a linear relation (considering the radio luminosity a proxy for the jet power, this is  equivalent to assume that the neutrino luminosity is proportional to the jet power). We are then able to connect the total radio flux of the FR0 population, $F_{\rm r,tot}$ to the total neutrino flux, $F_{\nu,{\rm tot}}$, i.e. $F_{\nu,{\rm tot}} = k F_{\rm r,tot}$. We note that, since the slope of the $N(F_{\rm r}$) distribution is less than 2, the total radio flux (and, by construction, the neutrino flux) received from FR0 is dominated by the sources at high fluxes. Therefore, even if below fluxes of 5 mJy we can still expect a large number of sources, the fraction of the total flux provided by them is small. In fact we have checked that using the full sample of ellipticals down to 1 mJy, the total flux increases only by a factor of 2. The all-sky total radio 1.4 GHz flux ($\nu F_{\nu}$) from FR0 can be derived summing the flux of the sources and multiplying by 4 (since the survey covers 1/4 of the sky). One obtains: $F_{\rm r,tot}\simeq 10^{-12}$ erg cm$^{-2}$ s$^{-1}$. The all sky, all flavors neutrino flux in the energy range 0.1-10 PeV can be estimated to be (see Righi et al. 2017) $F_{\nu,{\rm tot}}=2.3\times 10^{-9}$ erg cm$^{-2}$ s$^{-1}$. The ratio is then $k=F_{\nu,{\rm tot}} /F_{\rm r,tot}\simeq 2\times 10^{3}$. By construction, this parameter also links the neutrino and the radio luminosity for each source. Therefore, for a generic FR0 we can write: $L_{\nu}=2\times 10^{3} \times L_{\rm r}$ (where, again, for the radio luminosity we consider the $\nu L_{\nu}$ luminosity at 1.4 GHz).

This estimate of the ratio $L_{\nu}/ L_{\rm r}$ can be applied to the specific case of Tol 1326-379, for which we derived above an estimate of the jet radiative luminosity and power. The radio luminosity at 1.4 GHz is $\approx 10^{39}$ erg s$^{-1}$ (see Fig. \ref{sed}), from which we directly derive an estimate of the associated neutrino luminosity, $L_{\nu}\approx 10^{42}$ erg s$^{-1}$. This (see Table 1) is a fraction of about 1\% of the estimated jet power and about 10\% of the total radiative output (for the radiogalaxy case). Such an efficiency does not appear unreasonable (it is, for instance, much lower than that required in the blazar scenario, see Murase et al. 2014). With this luminosity the expected detection rate in IceCube in the $E_{\nu}\sim 0.1-1$ PeV range would be $N_{\nu}\approx \phi_{\nu} A_{\rm eff}$, where the flux $\phi _{\nu}=L_{\nu}/(4\pi d^2_{L}E_{\nu})\approx 2\times 10^{-14} $ cm$^{-2}$ s$^{-1}$ and $A_{\rm eff}\simeq 10^5$ cm$^2$, giving $N_{\nu}\approx 5\times 10^{-2}$ yr$^{-1}$, safely below the limit for multiple detections.

%Therefore, in the following estimate we assume $\delta\approx 1$.

In the following we discuss the possible neutrino production mechanisms acting in FR0. We note that for both $p\gamma$ and $pp$ mechanisms the average energy of the produced neutrinos is $E_{\nu }\approx E_{\rm p}/20$ and thus to produce neutrinos of energy $\sim$1 PeV through these reactions, protons with an energy $E_{\rm p}\approx 20 E_{\nu}=2\times 10^{16}$ eV are required. 

It is generally assumed that the acceleration timescale is related to some multiple $\eta$ of the gyration time (e.g. Rieger et al. 2007), $\tau_{\rm acc}\simeq 2\pi E_{\rm p}\eta /eBc$. The most relevant losses suffered by the protons in the jet are due to adiabatic expansion (as shown below the photomeson losses can be neglected), characterized by a  timescale $\tau_{\rm ad}\simeq R/c$. Using this upper limit for the loss timescale, the expected maximum energy of the accelerated protons, fixed by the condition that the acceleration time is equal to the cooling time, is $E_{\rm p, max} \approx 3\times 10^{17} B_{-1} R_{16} \, \eta^{-1}$ eV (see also Tavecchio 2014 and Tavecchio \& Ghisellini 2015), where we used physical parameters derived  above for the jet. Therefore, 100 PeV protons can be in principle accelerated in FR0 jets.

\subsection{Neutrinos from $p\gamma$ reactions}

A  first possibility to consider is that neutrinos are produced within the jet by accelerated protons interacting with the soft synchrotron radiation. The efficiency of the neutrino production through the photomeson reaction is measured by the factor $f_{p\gamma}=t_{\rm dyn}/t_{p\gamma}$, the ratio of the  dynamical timescale $\tau_{\rm dyn}\approx R/c$ and the cooling time of the protons through the  $p\gamma$ reaction, $\tau_{p\gamma}=[c n_{\gamma} <\sigma _{p\gamma}\kappa>]^{-1}$, where $n_{\gamma}$ is the number density of the photons above threshold and $<\sigma _{p\gamma}\kappa>\simeq 10^{-28}$ cm$^{2}$ is the $p\gamma$ cross section weighted by inelasticity (all these quantities are measured in the jet frame). The produced neutrino output, $L_{\nu}$ is of the order of $L_{\nu}\approx f_{p\gamma}Q_{\rm p}$, where $Q_{\rm p}$ is the proton injected power.

Protons with $E_{\rm p}\approx10^{16}$ eV interact with soft-UV photons with frequency higher than $\nu_{\rm s}\simeq 3\times 10^{15}$ Hz. Taking  Tol 1326-379 as benchmark, we can estimate that the luminosity at this frequency is $\nu_{\rm s}L(\nu_{\rm s})\approx 2\times 10^{42}$ erg s$^{-1}$. Assuming the parameters derived for the radiogalaxy case we found a (comoving) number density at this frequency $\nu_{\rm s}n_{\gamma}(\nu_{\rm s})\simeq 5\times 10^6$ cm$^{-3}$, implying a cooling time $t_{p\gamma}\simeq 1.2\times 10^{12}$ s. For the efficiency  we derive a quite small value, $f_{p\gamma}\simeq 10^{-6}$. Considering the value of the neutrino luminosity derived in the previous section, $L_{\nu}\approx 10^{42}$ erg s$^{-1}$, this small efficiency would imply a quite large power in the relativistic protons, $Q_{\rm p}\simeq L_{\nu} \, f_{p\gamma}^{-1}\simeq 10^{48}$ erg s$^{-1}$. Using the source parameters  obtained for the blazar case we find an analogous result. 

An extreme possibility to decrease the required proton power is to increase the soft photon density assuming a much compact emission region. Alternatively, one could assume an analogous of the spine-layer scenario used for BL Lac objects (Tavecchio et al. 2014). However, in both cases it seems hard to decrease $Q_{\rm p}$ by four orders of magnitude.

\subsection{Neutrinos from $pp$ reactions}

In a second scenario we postulate that the cosmic rays, accelerated within the jet at subpc scale, efficiently leave the flow at some larger distance (fixed by the condition that the magnetic fields decrease enough to allow them to escape). After leaving the jet, the protons can diffuse in the host galaxy, in which they interact with the relatively dense galactic medium through $pp$ reactions. In strict analogy with the scenario developed for star forming galaxies, we can estimate the efficiency of the neutrino emission $f_{pp}$ comparing the collisional energy loss timescale, $\tau_{\rm pp}$ and the residence time of the relativistic protons $\tau_{\rm res}$, $f_{pp}=\tau_{\rm res}/\tau_{pp}$.

The energy loss timescale can be written as:
\begin{equation}
\tau_{\rm pp}=\frac{1}{0.5 \, c \, \sigma_{\rm pp} n_{\rm p}}\simeq 3.5\times 10^7 \left( \frac{n_p}{\rm cm^{-3}}\right)^{-1} {\rm yr},
\label{taupp}
\end{equation}
where $\sigma_{pp}\approx 5\times 10^{-26}$ cm$^{-2}$ is the $pp$ cross section, 0.5 is the inelasticity and $n_{\rm p}$ is the density of the target cold protons (e.g. Loeb \& Waxman 2006).

The value of the residence time of the relativistic protons within the host galaxy is a parameter quite difficult to estimate. For illustration, we adopt the value derived for cosmic rays diffusing within our Galaxy. For protons of energy $E_{\rm p}$ the residence time can be written as:
\begin{equation}
\tau_{\rm res}(E_{\rm p})=3\times10^{7} \left( \frac{d}{\rm 1 \, kpc}\right) \left(\frac{E_{\rm p}}{10 \, {\rm GeV}}\right)^{-\alpha_d} {\rm yr},
\label{taures}
\end{equation}
where, in case of galactic cosmic rays, $\alpha_d=0.3-0.6$ (e.g. Blasi 2013) and $d$ is the size of the region containing the cosmic rays. 

%We estimate the required parameters in the case of the $pp$ reaction using the approximate relations provided by Becker Tjus et al. (2014). For a proton spectrum $N(E_{\rm p})=K_{\rm p}E_{\rm p}^{-2}$, the neutrino luminosity in the range 0.1-10 PeV can be expressed by:
%\begin{equation}
%L_{\nu}\approx 2.5 \, n_{\rm p} \, d \, \sigma_{pp} K_{\rm p}   	
%\end{equation}
%where $\sigma_{pp}\approx 3\times 10^{-26}$ cm$^{-2}$ is the $pp$ cross section, $n_{\rm p}$ is the density of the target cold protons, $d$ is the size of the region containing the targets. The product of the two latter quantities can be considered to express the column density of the target material, $N_{H}=n_{\rm p} \times d$. The normalization constant can be directly linked to the proton injected luminosity, $K_p=Q_{\rm p}/\ln(E_{\rm p,max}/E_{\rm p, min})$, in which the logarithmic term $\ln(E_{\rm p,max}/E_{\rm p, min})\approx 10$.
%Considering the neutrino luminosity derived above, for Tol 1326-379, we obtain: $Q_{\rm p}\simeq 10^{46} \, (N_H/10^{21} \, {\rm cm}^{-2})^{-1}$ erg s$^{-1}$.

For radiogalaxies, current estimates of the column density are around $N_H=10^{21}-10^{22}$  cm$^{-2}$ implying, for the cold gas, number densities in the central few kpc around $n_{\rm p}\sim 0.1-1$ cm$^{-3}$  (e.g. Balmaverde et al. 2006). For illustration we assume $d=3$ kpc, $n_p=1$ cm$^{-3}$, corresponding to $N_H\sim 10^{22}$ cm$^{-2}$. The loss timescales is therefore $\tau_{\rm pp}=3.5\times 10^{7}$ yr.  For $E_{\rm p}=10^{16}$ eV (corresponding to protons producing neutrinos of $E_{\nu}\approx E_{\rm p}/20=500$ TeV) we obtain $\tau_{\rm res} = 1.5\times 10^{6}$ yr  for $\alpha_d=0.3$ and  $\tau_{\rm res} = 2.2\times 10^4$ yr for $\alpha_d=0.6$, implying efficiencies in the range $f_{pp}\approx 6\times 10^{-4}-5\times10^{-2}$. With the most optimistic value $f_{pp}=0.05$, the required luminosity in relativistic protons for our benchmark case is therefore $Q_{\rm p}\simeq L_{\nu}f_{pp}^{-1}=2\times 10^{43}$ erg s$^{-1}$, ten times smaller than the jet power (Table 1). For $f_{\rm pp}=6\times 10^{-4}$ the required power would be $1.5\times 10^{45}$ erg s$^{-1}$ about ten times larger than $P_{\rm j}$ (this only accounts for cosmic rays at $E_{\rm p}\approx 10^{16}$ eV; the power increases by a factor $\approx$ 5-10 assuming a spectrum extending down to GeV energies with slope 2). Therefore, the required budget would be similar to or even less demanding than in the case of scenarios invoking other sources, e.g., blazars (e.g. Murase et al. 2014, Tavecchio et al. 2014).

Unfortunately, the estimate of $\tau_{\rm res}$ is subject to several uncertainties related to the properties of the magnetic fields associated to the host galaxies of FR0 and the related diffusion of ultra-relativistic protons. In fact, the magnetic field in elliptical galaxies (the typical host of jetted AGNs) are rather poorly characterized. The presence of substantial magnetic field turbulence in these galaxies (determining the effective scattering of the relativistic protons and thus a long residence time) is however advocated in e.g., Moss \& Shukurov (1996). It is also likely that the presence of the propagating jet can increase the average magnetic field and inject further turbulence in the intergalactic medium. 

\begin{figure}
\vspace*{-3 truecm}
\hspace*{-0.9 truecm}
\includegraphics[width=9.cm]{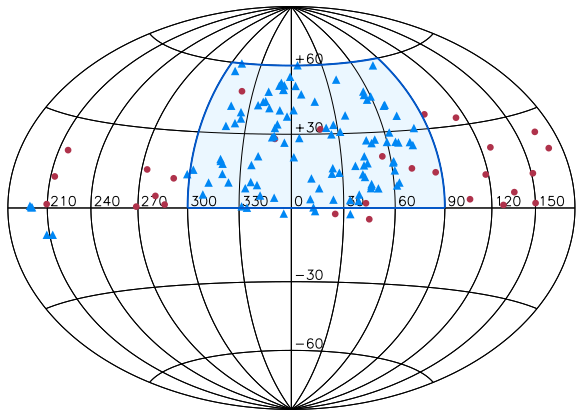}
\vspace*{-3truecm}
\caption{Sky map (in equatorial coordinates) showing the region containing the FR0 of the sample (light blue triangles) and the muon neutrinos from Aartsen et al. 2016 (red dots, for clarity we do not report the errors). The limited area containing the FR0 makes difficult to study a possible spatial correlation with the neutrino arrival directions.}
\label{map}
\end{figure}

\section{Discussion}

We have proposed that the emerging population of weak radio-galaxies dubbed ``FR0" can be the sources of (at least a fraction of) the neutrinos detected by IceCube (we note that Jacobsen et al. 2015 also briefly mentioned this possibility). We have modelled the SED of the gamma-ray emitting FR0 Tol 1326-379 in the framework of the one-zone leptonic model, deriving the values of the basic physical parameters of the jet. The energy budget associated to the relativistic protons required for each FR0 appears too large if the neutrino emission occurs through the $p\gamma$ channel, as required by the widely assumed scenario for neutrino production in jets. 

An alternative is based on the assumption that the high-energy protons accelerated in the jet are able to efficiently escape in the host galaxy, where they can loose energy through the $pp$ channel, scattering off the intergalactic medium -- in analogy with the situation occurring in star forming galaxies (in that case cosmic rays are injected by expanding supenova remnants) or as envisioned in the AGN-driven model (Lamastra et al. 2016). In this scenario the required energy budget could be acceptable and the required power to sustain the neutrino flux could be smaller than the inferred jet power. A higher efficiency of the neutrino emission could be foreseen if the cosmic rays escaping from the host can still scatter in the gas associated to the group/cluster possibly embedding the FR0 (K. Murase, priv. Comm.)
A more detailed discussion is currently hindered by the poor knowledge of the cosmic ray transport and confinement in the host galaxies of FR0. We note that  from $pp$ reactions we expect a comparable amount of energy in the form of $\gamma$-rays injected at $0.1-1$ PeV. These photons promptly trigger electromagnetic cascades interacting with CMB photons (a PeV photon has a mean free path comparable to the radius of the galaxy). For low-redshift sources as FR0s, the resulting spectrum is hard, $F_{\rm rep}\propto E^{-0.5}$ (Berezinsky \& Kalashev 2016) and potentially detectable only above several tens of TeV.

For illustration we show in Fig. \ref{map} the positions of the 108 FR0 radiogalaxies (blue triangles) and the the arrival directions of muon neutrinos (i.e. those with the best reconstructed direction) from Aartsen et al. (2016). The large majority of the FR0 are concentrated in the light blue area. Clearly the sparse number of neutrinos and the relatively small region covered by the FR0 sample do not allow the study of the positional correlations as, for instance, in the analysis performed by Palladino \& Vissani (2017) for the high energy emitting BL Lac. Larger survey looking for FR0 galaxies are clearly required to perform such kind of studies.

We finally note that our analysis is based on the assumption that a unique class of objects (FR0 in our case) provides the bulk of the neutrino background. However, it is possible (or even likely) that the observed flux is the result of the cumulative contribution of several classes of sources, in analogy with the $\gamma$-ray background. A mixed composition of neutrino emitters could also reconcile the association of the recent IceCube event with the BL Lac TXS 0506+056 and the constraints derived by Palladino \& Vissani (2017), which seems to rule out a dominant contribution from BL Lac sources.
%This would decrease the energetic demand derived here. 
In this view, all types of radiogalaxies could emit neutrinos (e.g., Becker-Tjus et al. 2014, Hooper 2016). Present evidence suggests that FR0s greatly outnumber FRIs, implying that, even taking into account the larger power of FRI jets, FR0s are expected to dominate the neutrino output from radiogalaxies. 

\section*{Acknowledgments}
We thank K. Murase for comments. The authors acknowledge contribution from a grant PRIN--INAF--2014 and the grant INAF CTA--SKA. Part of this work is based on archival data and  on--line services provided by the ASI SDC.

\end{document}